\documentclass[a4paper,11pt]{article}
\pdfoutput=1 

\usepackage{jheppub} 

\usepackage[T1]{fontenc} 

\pdfoutput=1 

\usepackage[T1]{fontenc} 
\usepackage{color}   
\usepackage{graphicx,float}
\usepackage{dcolumn}
\usepackage{bm}
\usepackage{slashed}
\usepackage{float}
\usepackage{amsmath}
\usepackage{amssymb}

\def\bb    #1{\hbox{\boldmath${#1}$}}

\def\2d{{{}_{\rm 2D}}}         
\def\4d{{{}_{\rm 4D}}}         

\title{\boldmath  The Open Relativistic Two-body Problem}

\author{A.V.Koshelkin}
\affiliation{National Research Nuclear University MEPhI, \\ 
Kashirskoye shosse, 31, 115409 Moscow,\\ Russia}

\emailAdd{and.kosh59@gmail.com}

\abstract{The open relativistic two-body problem, when two interacting particles also are in external potentials, is considered  in terms of  the principle of the least action. Based on the consistent modification of   the relativistic version Newton's third law in external fields,  the exact covariant operator equations,  which govern dynamics of either a scalar particle or each of the components of the 16 component spinor of spin-$1/2$ fermions, depending on the particle type, are derived  in the center-mass and relative motion variables, beyond the consideration in the Breit frame only. The class of external fields and interaction potentials,  when the two-body problem can be covariantly reformulated in (3+1) phase space of relative motion variables, uncoupled from  the center-mass motion of such a system, is outlined. In the case of fermions the new $\gamma$-matrix basis generating the Dirac-like equation for the 16 component spinors in the (3+1) phase space is found. The chosen basis allows us to decouple the derived Dirac-like equation into four independent equations governing the dynamics of the four-component spinors. The developed approach is examined  in studying the low-energy positronium states in  magnetic fields.}

\begin{document} 
\maketitle
\flushbottom

\section{Introduction}
\label{intro}

The relativistic two-body problem, which has actively studied for more   than forty years, beginning from 80th \cite{Teo71,Teo76,Droz,Kal,Teo78,Kom,Are,Kir,Dom78,Dom80,Sad86-1,Sad86-2,Sad94,Sad89,Cra84,Cra87,Cra90,Cra01,Cra03,Cra06,Cra10,Bar87,Bar91,Log}, is of  great interest now in connection primarily  with the indirect searches of   the dark matter\cite{Agu,Perez,Jung,Ber,Gel}. It also  concerns researches when   strong magnetic fields play an important role in forming signals from both the astrophysics  objects\cite{Bro,Bus,Dun,Hard06} and the  strong interacting matter, arising in non-central collisions of high energy particles\cite{Fuku,Vor,Tuc}.  The value of the magnetic field strength in the latter case is estimated to be about $B\sim 10^{18}- 10^{19} G$ at RHIC and LHC energies\cite{Fuku,Vor,Tuc} what is of the order  and more than  the squared pion mass. Besides that, magnetic fields play a key role in  forming signals from  the interior area of neutron stars\cite{Dun,Hard06}  where  the magnetic field  strength turns out to be   of the same   order of magnitude.

The first attempt to describe the relativistic two-body system was made by summing two free Dirac equations,  adding to them interaction potential \cite{Brei29,Brei30}, that resulted in the relativistic non-invariant Breit equation.
On other hand,  the manifestly covariant    Bethe-Salpeter equation\cite{BSE}, which is  suitable for solving such kind   problems,  was found to have\cite{Nak} the negative normalized  solutions in the case  of  two one half spin particles. The lack of the approaches mentioned above can  be eliminated with Dirac's Hamiltonian constraint technique\cite{Dir,Ber}, decreasing a number of the Hamiltonian variables by means of establishing relations between them. It is achieved  due to  the Poincare invariant functions containing interaction terms $\Phi_i (p_1, m_1; p_2, m_2, q_{12}...)$

\begin{eqnarray}\label{eq0}
 p^2_i - m^2_i =\Phi_i (p_1, m_1; p_2, m_2, q_{12},...), 
\end{eqnarray}
where $i=1,2$ numerates particles, whereas the functions  $\Phi_i$ depends on  the particle momenta  $p_i$ and on  the parameters $q_{12}$ characterizing interaction between particles.

On the other hand,   in non-relativistic physics  the III-rd Newton law restricts  a set of  the interaction and external fields which allow to reduce the two-body problem to the single particle dynamics. The similar restrictions dictated  by the relativistic version of the III-rd Newton law to   external  and interaction  fields should obviously take place in the relativistic case. The relativistic version of this law is given by a formula\cite{Cra84,Cra87,Cra90,Cra01}

\begin{eqnarray} \label{eq000} 
&& P p = 0, 
\end{eqnarray}
where$P$ and $p$ are the total and relative momenta of particles. Such a restriction turns out to b the necessary condition of the compatibility of equation in studying of the two-body problem in the constraint dynamics approach\cite{Dir,Ber}.

Based on the  constraint dynamics ideas\cite{Dir,Ber} the classical relativistic two-particle system has been  studied \cite{Droz,Kal,Teo78,Kom} for the first time, where a consistent and covariant approach has been developed for the relativistic canonical Hamiltonian mechanics. 
For more than forty years, beginning from 80th, the constraint dynamic approach   have been  actively and successfully developed in studying the quantum two-body systems,  and applied for describing real compound systems in QED and QCD \cite{Sad86-1,Sad86-2,Sad89,Cra84,Cra87,Cra90,Cra01,Cra03,Cra10}. Following the constraint dynamics  the two interacting particle systems, involving spin-0 bosons and/or spin-(1/2) fermions, is studied \cite{Sad86-1,Sad86-2}. The obtained results and is applied to the description of the quarkonium states \cite{Sad86-2}. In the framework of the constraint dynamics approach  the $4\times 4$ matrix wave function satisfying  an equation of the Pauli-Schrodinger type is constructed\cite{Sad94}. Later, the developed  approach   was  generalized to the case of the N interaction particle\cite{Sad89}.

Along  with researches\cite{Sad86-1,Sad86-2,Sad89},  by developing previous results\cite{Teo76,Teo78},  the quantum consideration of two spinless\cite{Cra84} and spin-$1/2$\cite{Cra84} particles under  vector and scalar  interaction, have been developed, using the  constraint mechanics for the  two-body Klein-Gordon  and Dirac  equations. The hyperbolic interaction structure of these equations,   established earlier\cite{Cra84,Cra87},  allows\cite{Cra90}  us  to include the pseudoscalar, pseudovector, and tensor interactions in the co responding  motion  equations.  The Hamiltonian formulation of the relativistic N-body problem in a separable two-body basis, in which the particles interact pair-wise through scalar and vector interactions, is developed in Ref.\cite{Cra01}.  The worked out  approaches\cite{Cra84,Cra87,Cra90,Cra01} have been  further  applied to researches  in a meson spectroscopy \cite{Cra06,Cra10}, and in  nuclear-nuclear collisions\cite{Cra03}. 
  The two-body states in the  case of  QED interaction\cite{Bar87} and when the interaction is realized due to  a  neutral boson\cite{Bar91}, has been studied in terms of the action integral in the coordinate representation  that leads to the non-covariant Breit-like equation in the coordinate space-time.

The dynamics of the two-body systems is studied  in the Lagrange formalism in the case when such a system consists of two classical relativistic  particles\cite{Dom78,Dom80}. 
The singular-Lagrangian approach  to the two-body problem is derived therein\cite{Dom78,Dom80}. 

The relativistic two-particle systems in external fields which consist of scalar\cite{Droz00,Droz09} and fermion\cite{Bij00} particles are considered in the context of the covariant separation of the relative and mutual motion such systems. Having been mainly  concentrated on the mutual motion and on constraint relations concerning  external field\cite{Droz00,Bij00,Droz09},  the reconstruction of  the relative motion spectrum as well as the spin mixing effect   turn out beyond the consideration mentioned above\cite{Droz00,Bij00,Droz09}.

Despite the active use of the relativistic version of the third Newton law\cite{Cra84,Cra87,Cra90,Cra01}, the requirements for the interaction  potentials that allow to say about the two-body problem in the relativistic case, as well as the modification of the third Newton law in external field potentials,  have still remained to be undiscussable. The same concerns arising the special phase space for  the two-body system  evolution as well as  relations between the seed constituent and effective phenomenological potential what are beyond previous considerations\cite{Sad86-1,Sad86-2,Sad89,Cra84,Cra87,Cra90,Cra01,Cra03,Cra10}.
Besides that, even in the absence of external field, the  Dirac equation  derived in earlier\cite{Sad86-1,Sad86-2,Sad89,Cra84,Cra87,Cra90,Cra01,Cra03,Cra10} turn out to be very complicate for  analyzing the dynamics of  fields  in real physical situations. Therefore, the question whether it is possible to find some new relevant $\gamma$-matrix basis which allows to simplify the Dirac equation structure in the two-body problem, has  remained unsolved.

In the present paper the two-body problem in the external vector fields is studied in the Lagrange formalism beyond the constraint Hamiltonian dynamics.   The exact covariant operator equations governing scalar particles and spin-$1/2$ fermions have been  derived. The obtained equation are formulated in the standard space-time variables of  the relative and  center mass motion of the two-body system, beyond the consideration in the Breit frame of reference only\cite{Sad86-1,Sad86-2,Sad89,Cra84,Cra87,Cra90,Cra01,Cra03,Cra10}.  The class of  interaction and external field potentials when the two-body problem can be covariantly reformulated in (3+1) phase space of the relative motion variables, which are uncoupled from the center mass variables,  is pointed out. 

Provided that Newton's third law occurs,  the Klein-Gordon equation, which  governs either dynamics of  a scalar particle or describes each of a  single component of the 16 components spinor of the spin-$1/2$ fermion, depending on the particle type,  has been derived  in the (3+1) phase space.  In the case of $1/2$-spin particles we find a new $\gamma$-matrix basis which allows us  to go to the Dirac-like equation  which is decoupled  in four independent equations for four different components of the  initial 16 component spinor.  In terms of  the derived Dirac-like equations the  Lagrangian  governing the dynamics of two spin-$1/2$ fermions in an external field has been found. This  Lagrangian generates the  conservative  fermion  current.

  In the case of pure  QED the developed approach is applied to  studying  the low energy  states of a positronium in a magnetic field, taking into account spin-mixing effect.  It is shown that strong magnetic fields result in essentially reconstructing the spectrum of positronium states and decreasing the life-time of the positronium ground state.
 
 The paper is organized as follows.  The system consisting of  two interacting scalar particles in external fields is studied in Section 2.  Two interacting particles with spin-$1/2$ in external fields are considered in Section 3.  Section 4  is devoted to  studying the low energy  states of a positronium in a strong uniform  magnetic field.  The last section is the conclusion. The details of some  calculations are given in an appendix.
 
\section{Two interacting spinless particles in external fields}
\label{sec:1}
We consider two scalar particles with   masses $m_1$ and $m_2$ which interact  by means of  vector $A_{int}^\nu (x)$ and  scalar $S_{int}(x)$ potentials, and which also are in an external vector $A_{ext}^\nu (x)$ and scalar $S_{ext}(x)$  potentials. Such external potentials  transform the standard confined two-body problem into the open one. The action integral of such a system of particles consists of two terms

\begin{subequations}
\begin{eqnarray}\label{eq1}
&&{\frak A} =\sum\limits_{i=1}^2 \frac{\lambda_i}{2} \int dx_i\Big(  |  \pi^\mu_{i} \Phi_i (x_i)|^2  - \Phi^\ast_i (x_i) ( m_i + {\cal S}_i(x_{i}))^2    \Phi_i (x_i ) \Big)
\\
&& \pi^\nu_{i} \equiv  p^\nu_{i} + {\cal A}_{i}^{\nu} (x_{i})= i\partial^\nu_{i} + {\cal A}_{i}^{\nu} (x_{i}),
\end{eqnarray}\label{eq1-a}
\end{subequations}
where $i=1,2$ enumerates  particles,  $ \Phi_{i} (x_{i}) $ denotes a  scalar field, ,  $p^\nu_i = i\partial^\nu_i = (i\partial^0,\; -i\nabla)$ is a particle momentum; $ g_{\nu \mu}$ is the metric tensor which diagonal is $diag= (1,-1,-1,-1)$, whereas $x=(x^0, \bb x)$. The parameters $\lambda_{i}$ are the Lagrange multipliers; ${\cal A}_{i}^\nu (x_{i})$ and ${\cal S}_{i} (x_{i})$ are the vector and scalar potentials acting on the field  $ \Phi_{i} (x_{i}) $, respectively.  These potentials consist of two terms which describe  interaction between particles inside the two-body and an external fields acting on each particle independently.
For brevity,  the coupling constants  are not written explicitly, and are included into the potential ${\cal A}_{i}^\nu (x_{i})$ and ${\cal S}_{i} (x_{i})$.

Further consideration requires   to discuss the structure of the interaction terms in Eq.(\ref{eq1}). The fields governing interaction inside the confined two-body problem have to obviously take into account the states of both interaction particle, whereas any external field acting on the each of the considered particles is the same, differing by the coupling constant in the interaction term. Then, we write the interaction potentials as follows

\begin{subequations}
\begin{eqnarray}\label{eq8}
&& { \cal A}^\nu_{i} (x_{i}) = { A}^\nu_{i } (x_{i}) + \frac{  \int dx_j  \Phi^\ast_{j} (x_j) { A}^\nu_{ ij } (x_{i}-x_j )\Phi_{j} (x_j)   }{ \int dx_j \Phi^\ast_{j} (x_j)\Phi_{j} (x_j) }, \nonumber \\ \\ \label{eq8a}
&&{ \cal S}_{i} (x_{i}) = { S}_{ i } (x_{i}) + \frac{  \int dx_j  \Phi^\ast_{j} (x_j) { S}_{ ij } (x_{i}-x_j )\Phi_{j} (x)   }{ \int dx_j \Phi^\ast_{j} (x_j)\Phi_{j} (x_j) }.
\end{eqnarray}
\end{subequations}
 
Here ${ A}^\nu_{i} $ and $ S_{j}$ denote  the potentials  affecting  the $i$-th particle by an external field, whereas ${ A}^\nu_{ij} $ and $ S_{ij}$ supplied by a double subscribe notate  the potentials acting by  the $j$-th particle on the $i$-the one,  inside the two-body system ( $i,j = 1,2, i\ne j$).  
All of ${ A}^\nu_{ij} $ depends , in general,  on the fields  $\Phi_{j} (x_j) $. However, we ignore such a dependence, restricting by consideration of the so-called mean field approximation  for the two-particle interaction.

As for the  external potentials, to eliminate interaction between external and interaction potential we demand them orthogonality which means the following  in the case of the vector potentials

\begin{eqnarray}\label{ortho}
&&{ A}^\nu_{ij}  ( A_\nu)_{k}  =0, ~~~i,j,k = 1,2.
\end{eqnarray}
The restrictions to the external scalar potentials will be discussed below in Sec.2.4.

\subsection{Free particles}

Let us consider the two-body problem in case of free particle to clarify selection of the Lagrange
multiplier $\lambda_{1,2}$.

We firstly take the Lagrange multipliers to be 

\begin{eqnarray}\label{eee}
&&\lambda_{1} =\Big( \int dx_2 \Phi^\ast_{2} (x_2)\Phi_{2} (x_2)\Big),~~~\lambda_{2} =\Big( \int dx_1 \Phi^\ast_{1} (x_1)\Phi_{1} (x_1)\Big).\nonumber \\
\end{eqnarray}

Then, we   introduce the  canonically conjugation  variables\footnote{The same momenta are used earlier\cite{Teo71,Teo76}}

 \begin{subequations}\label{eq11}
\begin{eqnarray}\label{eq11-1}
&&~P=p_1+p_2, ~~~p=\mu_1 p_2 - \mu_2 p_1,  \\
&&p_1 = \mu_1 P - p, ~~~p_2 = \mu_2 P + p, \\
&&X=\frac{1}{2}(x_1 +x_2), ~~~~x =x_2-x_1.\label{eq11-2}
\end{eqnarray}
\end{subequations}
where 
\begin{subequations}
 \begin{eqnarray}\label{eq11a-1}
&&\mu_1 = \frac{1}{2}\left(1+ \frac{m_1^2 -m_2^2 }{M^2}\right), \mu_2 =\frac{1}{2}\left(1- \frac{m_1^2 -m_2^2 }{M^2}\right),  \nonumber \\ \\
&&\mu_1 +\mu_2 = 1, ~~ ~~ M= |p_1+p_2 | = |P|.
\end{eqnarray}
\end{subequations}
In Eq. (\ref{eq11-1})  $P$  and $p$ are the total and relative  momentum of the two-body system, respectively, while  $M=|P|=|p_1+p_2|$ is its invariant mass.  The momenta  $P,p,$ and coordinates $X,x$ satisfy the following commutative relations

\begin{eqnarray}\label{commut}
&&[P_j,x_k]= [p_j,X_k]=0,\nonumber \\
&& [P_j,X_k]=-i\delta_{jk},[p_j,x_k]=-i\delta_{jk}, ~j,k=1,2,3.
\end{eqnarray}

In terms of such new  variables  the action integral becomes to be equal to 

\begin{eqnarray}\label{eq14}
&&{\frak A}_+ =  \int   dX dx \int   dX dx\Phi^\ast  (X, x)\big(p^2+  E_w^2 - m_w^2- \frac{m_1^2 -m_2^2 }{M^2}Pp\big)  \Phi (X, x),    
\end{eqnarray}
where we introduce an effective energy and an effective mass of the two-body  system\cite{Teo76,Teo78,Cra84,Cra87} , 

\begin{eqnarray}\label{eq18aa}
&&E_w= \frac{1}{2M} (M^2- (m_1^2 +m_2^2 ) ), ~~~m_w =\frac{ m_1 m_2   }{M}, 
\end{eqnarray}
and the two-body field $\Phi (X, x)$

\begin{eqnarray}\label{eq18aab}
&&  \Phi (X, x) =  \Phi_1  (X-x/2 )\Phi_2 (X+x/2).
\end{eqnarray}
Varying with respect to the field $\Phi^\ast  (X, x)$ we obtain the motion equation

\begin{eqnarray}\label{eq14me}
&&\bigg( { E}_w^2 + p^2-{  m}_w^2- \big({ {\mu_1} }-{ {\mu_2} }\big){ P}{ p}\bigg)\Phi (X, x) =0.
\end{eqnarray}

Let us  take the Lagrange multiplier as follows

\begin{eqnarray}\label{eee-}
&&\lambda_{1} =\Big( \int dx \Phi^\ast_{2} (x)\Phi_{2} (x)\Big),\lambda_{2} =-\Big( \int dx \Phi^\ast_{1} (x)\Phi_{1} (x)\Big).\nonumber \\
\end{eqnarray}
Then, we get the action integral ${\frak A}_-$

\begin{eqnarray}\label{eq14}
&&{\frak A}_- =  \int   dX dx \int   dX dx\Phi^\ast  (X, x)\big(  Pp\big)  \Phi (X, x).    
\end{eqnarray}

Such an action integral generates the motion equation

\begin{eqnarray}\label{eq14me-}
&&({ P}{ p})~\Phi (X, x) =0.
\end{eqnarray}

This motion equation corresponds to the relativistic version of the third Newton's law\cite{Cra84,Cra87} and is in entire relation to Eq.(19)\cite{Cra87}. Such a results is physically understandable since the Lagrange multipliers selection dictated by Eq.(\ref{eee-})  keeps the particle interaction only, eliminating their kinematics.

Provided that Eq.(\ref{eq14me-}) the motion equation takes a form

\begin{eqnarray}\label{eq14me-11}
&&\bigg( { E}_w^2 + p^2-{  m}_w^2\bigg)\Phi (X, x) =0.
\end{eqnarray}

This equation is covariant in ${\frak R}^3_1 (X) \otimes {\frak R}^3_1(x)$ space corresponding to the  both  center-mass and relative motion. Such  types of motion can be obviously decoupled looking for the solution of Eq.(\ref{eq14me-11}) as

\begin{eqnarray}\label{eq14me11--}
\Phi (X, x) =\chi (X) \varphi ( x), 
\end{eqnarray}

where  $\chi (X)$ is the eigenfuncion of an equation
\begin{eqnarray}\label{eq14me--}
&& { E}_w^2 \chi (X) = { E}_w^2 (P)  \chi (X).
\end{eqnarray}

As a result, we go to the Eq.(\ref{eq14me-1}) for the function $\phi ( x)$  where ${ E}_w^2 (P)$ has already been a $C$-function. 

\begin{eqnarray}\label{eq14me-1}
&&\bigg( { E}_w^2 (P) + p^2-{  m}_w^2\bigg)\varphi ( x) =0.
\end{eqnarray}
Although this equation is covariant in the ${\frak R}^3_1(x)$ but it cannot be reduced to the standard Klein-Gordon equation. 

We look for the solution of Eq.(\ref{eq14me-1}) as follows

\begin{eqnarray}\label{eq14me-1a}
&&\varphi ( x) = \varphi ( \bb x) \exp(-i\tau p^0).
\end{eqnarray}

We breaking the   ${\frak R}^3_1(x)$ covariance in favor of keeping  the  Klein-Gordon form of the motion equation, that turns out to be reasonable under  studying two interaction particles,  and  introduce $\varepsilon_w $ and $ \theta_w$ 

\begin{subequations}\label{varep}
\begin{eqnarray}
&& { E}_w (P) =\varepsilon_w \cos \theta_w \\
&& p^0=\varepsilon_w \sin \theta_w .
\end{eqnarray}
\end{subequations}
As a result, we get 

\begin{eqnarray}\label{eq14me-2}
&&\bigg( { \varepsilon}_w^2+ \bb p^2-{  m}_w^2\bigg)\varphi (\bb  x) =0, 
\end{eqnarray}
the obtained equation  is the Klein-Gordon equation which is covariant in the ${\frak R}^3_1(\varepsilon_w , \bb x)$ phase space\cite{Teo71,Teo76,Droz,Kal,Teo78,Kom,Are,Kir,Dom78,Dom80,Sad86-1,Sad86-2,Sad94,Sad89,Cra84,Cra87,Cra90,Cra01,Cra03,Cra06,Cra10} rather than the standard coordinate  ${\frak R}^3_1(x)$ space-time.

\subsection{Interacting  particles}

In the case of interacting particles we construct the action integrals ${\frak A}_+$ and ${\frak A}_-$ by taking the Lagrange multipliers $\lambda_{1,2}$ according to  Eqs.(\ref{eee}), (\ref{eee-}) as it has been done above for the free particles. Such action integrals generate the motion equations after going to the center mass  and  relative motion variables (see Eq.(\ref{eq11})).

\begin{eqnarray}\label{eq14me-int-1}
&&\bigg\{2(p^2  +  E_w^2 - m_w^2- (\mu_1 -\mu_2)Pp) + {\cal A}_1^2 (X-x/2) +2 {\cal A}_1 (X-x/2)(\mu_2 P -p) +\nonumber \\
&&  [ (\mu_2 P -p) ; {\cal A}_1 (X-x/2)] +  {\cal A}_2^2 (X+x/2) +2 {\cal A}_2 (X+x/2)(\mu_1 P +p)+
\nonumber \\
&&\  [ (\mu_1 P +p) ; {\cal A}_2 (X+x/2)]-2m_1 {\cal S}_1   (X-x/2) -{\cal S}^2_1   (X-x/2)+\nonumber \\
&& -2m_2 {\cal S}_2   (X+x/2) -{\cal S}^2_2   (X+x/2)\bigg\}\Phi (X, x)=0,
\end{eqnarray}

\begin{eqnarray}\label{eq14me-int-2}
&&-2Pp~\Phi (X, x) =  \bigg\{ {\cal A}_1^2 (X-x/2) +2 {\cal A}_1 (X-x/2)(\mu_2 P -p) +[ (\mu_2 P -p) ; {\cal A}_1 (X-x/2)] -\nonumber \\
&&  {\cal A}_2^2 (X+x/2) -2 {\cal A}_2 (X+x/2)(\mu_1 P +p)+
  [ (\mu_1 P +p) ; {\cal A}_2 (X+x/2)]-\nonumber \\
&&2m_1 {\cal S}_1   (X-x/2) -{\cal S}^2_1   (X-x/2)-2m_2 {\cal S}_2   (X+x/2) -{\cal S}^2_2   (X+x/2)\bigg\}\Phi (X, x),
\end{eqnarray}
where the square brackets denote, as usual, a commutator.

Obtained above Eq.(\ref{eq14me-int-2}) governs breaking  the 3-rd Newton law which naturally  takes place when there is no restriction to interaction potentials. 

Substitution Eq.(\ref{eq14me-int-1}) into Eq.(\ref{eq14me-int-2}) leads to

\begin{eqnarray}\label{eq14me-int-3}
&&\bigg\{(E_w^2+p^2 - m_w^2) + {\cal A}_1^2 (X-x/2)\mu_1 +2 {\cal A}_1 (X-x/2)(\mu_2 P -p)\mu_1 +\nonumber \\
&&  [ (\mu_2 P -p) ; {\cal A}_1 (X-x/2)] \mu_1+  {\cal A}_2^2 (X+x/2)\mu_2 +2 {\cal A}_2 (X+x/2)(\mu_1 P +p)\mu_2+
\nonumber \\
&&[ (\mu_1 P +p) ; {\cal A}_2 (X+x/2)]\mu_2-2m_1 {\cal S}_1   (X-x/2) \mu_1 -{\cal S}^2_1   (X-x/2)\mu_1+\nonumber \\
&& -2m_2  {\cal S}_2   (X+x/2)\mu_2 -{\cal S}^2_2   (X+x/2)\mu_2\bigg\}\Phi (X, x)=0.
\end{eqnarray}
 
Note, that  Eq.(\ref{eq14me-int-3}) can be also derived by summing and subtracting  the motion equation for each of the particles consisting of the two-body system.

\subsection{The two-body problem in the absence of external fields}

The center-mass and relative motion variables turn out to be uncoupled in Eq.(\ref{eq14me-int-3}) due external potentials since these potentials  depend on the coordinate each a particle independently (see Eqs.(\ref{eq8}), (\ref{eq8})).  We firstly consider  the situation when  there is no external potentials acting on the two-body system. We also assume that the potentials ${\cal A}^\mu_i  =( {\cal A}^0_1 , \bb  {\cal \bb A}_i ) ={\cal A}^\mu_i (\bb x) $ and ${\cal S}_i={\cal S}_i(\bb x) $ , where $i=1,2 $ numerates particles, depend only on the spatial relative coordinate $\bb x$. Moreover, let the following symmetry relation be

\begin{subequations}
\begin{eqnarray}\label{restric}
&& {\cal \bb A}^\mu_1 (\bb x) ={\cal \bb A}^\mu_2 (\bb x)\equiv {\cal \bb A}^\mu (\bb x), \\
&& {\cal A}^0_i (\bb x) =  {\cal A}^0_i (-\bb x), ~~~{\cal \bb  A} (\bb x) =-  {\cal \bb A} (-\bb x)\\
&& {\cal S}_i (\bb x) = {\cal S}_i (-\bb x).
\end{eqnarray}\label{restric-1}
\end{subequations}
Then,  we separate the center-mass and relative motion variables, following Eqs.(\ref{eq14me11--}), (\ref{eq14me--}). As a result,  in terms of $\varepsilon_w$ and $\theta_w$ given by Eq.(\ref{varep}) we obtain from  Eq.(\ref{eq14me-int-3})

\begin{eqnarray}\label{eq14me-int-4}
&&\bigg\{(\varepsilon_w^2+(\bb p +{\cal \bb  A} (\bb x))^2 - m_w^2) +\mu_1 ({\cal A}^0_1)^2 (\bb x) +\mu_2 ({\cal A}^0_2)^2 (\bb x) +2\mu_1 \mu_2  P^0({\cal A}^0_1 (\bb x)+{\cal A}^0_2 (\bb x)) 
\nonumber \\
&&+2  p^0(\mu_2 {\cal A}^0_2 (\bb x)-\mu_1{\cal A}^0_1 (\bb x)) -\mu_1{\cal S}^2_1 --\mu_2{\cal S}^2_2   (\bb x)-2(m_1 \mu_1 {\cal S}_1 (\bb x)  +m_2 \mu_2 {\cal S}_2 (\bb x)  )   \bigg\}\varphi (\bb x)=0.\nonumber \\
\end{eqnarray}

We write the potentials ${\cal A}^0_i (\bb x)$ and $ {\cal S}_1 (\bb x)$ as follows

\begin{subequations}\label{pot-AS}
\begin{eqnarray}
&& {\cal  A}^0_1 (\bb x) =\frac{1}{\sqrt \mu_1 } {\cal  A}^0  (\bb x) \cos \theta_A, ~~ {\cal  A}^0_2 (\bb x) =\frac{1}{\sqrt \mu_2 } {\cal  A}^0  (\bb x) \sin \theta_A, \\
&&{\cal S}_1 (\bb x) =\frac{1}{\sqrt \mu_1 } {\cal S}^0  (\bb x) \cos \theta_S, ~~ {\cal S}_2 (\bb x) =\frac{1}{\sqrt \mu_2 } {\cal S}^0  (\bb x) \sin \theta_S.
\end{eqnarray}
\end{subequations}

Substituting Eq.(\ref{pot-AS}) into Eq.(\ref{eq14me-int-4}) we derive

\begin{eqnarray}\label{eq14me-int-6}
&&\bigg\{\varepsilon_w^2+(\bb p +{\cal \bb  A} (\bb x))^2 - m_w^2 +({\cal A}^0)^2 (\bb x) +2  {\cal A}^0 (\bb x)  P^0 \cos ( \theta_A -  \psi_A ) - 2 {\cal A}^0 (\bb x) p^0\cos ( \theta_A +  \psi_A )
\nonumber \\
&& -{\cal S}^2   (\bb x)-2{\cal S} (\bb x) (m_1 \sqrt \mu_1 \cos \theta_S   +m_2 \sqrt \mu_2 \sin \theta_S  )   \bigg\}\varphi (\bb x)=0,
\end{eqnarray}
where $\tan \psi_A = \sqrt {\mu_2 / \mu_1}$.

We take $\theta_A$ and $\theta_S$ so that the following equations takes place

\begin{eqnarray}\label{constr-1}
&&   P^0 \cos ( \theta_A -  \psi_A ) - p^0\cos ( \theta_A +  \psi_A )= \varepsilon_w
\end{eqnarray}

\begin{eqnarray}\label{constr-2}
 m_1 \sqrt \mu_1 \cos \theta_S   +m_2 \sqrt \mu_2 \sin \theta_S  =m_w.
\end{eqnarray}
Then, the motion equation takes a form

\begin{eqnarray}\label{eq14me-int-7}
&&\bigg\{(\varepsilon_w + {\cal A}^0 (\bb x))^2+(\bb p +{\cal \bb  A} (\bb x))^2 - (m_w +{\cal S} (\bb x))^2\bigg\}\varphi (\bb x)=0. 
\end{eqnarray}
If the solution of Eq.(\ref{eq14me-int-7}) is known the general solution of the problem can be  written using Eqs.(\ref{eq14me--}), (\ref{eq14me-1a}), provided that Eq.(\ref{restric-1}) also takes place.

In the Breit frame, when $\bb p_1 + \bb p_2 =0 $ the obtained equation Eq.(\ref{eq14me-int-7}) coincides with the motion equations have derived earlier\cite{Cra84,Cra87}.

\subsection{The two-body problem in external fields}

We consider the open two-body problem when the two-particle interaction system is also in external fields. The possibility to factorize   the center-mass and relative motion variables, as it has been done above in the absence of external potentials, is entirely determined  on the structure of the first terms in Eq.(\ref{eq8}), (\ref{eq8a}).   

In order to the center-mass and relative motion variables could be  separated the sums of external potentials ${ A}^\nu_{1 } (X-x/2 )+ { A}^\nu_{2 } (X+x/2 ) $ and ${ S}_{1 } (X-x/2 )+ { S}_{2 } (X+x/2 ) $  should depend only on either $X$ or $x$. Since such  a demand is impossible for the scalar field we exclude the external scalar field from our  consideration below , setting   ${ S}_{1 } (X-x/2 )= { S}_{2 } (X+x/2 ) =0$. As for the external vector fields   the center-mass and relative motion variables  are decoupled  when  the vector potentials are a linear function of coordinates

\begin{eqnarray}\label{factorizing}
&&{ A}^\nu_{1 } (X-x/2 )+ { A}^\nu_{2 } (X+x/2 ) = e^{\nu \mu  \alpha \beta} F_{\mu \alpha} (x_{ 1} \pm x_{ 2})_\beta, 
\end{eqnarray}
where $F_{\mu \alpha}$ is some 4-tensor being independent on coordinates. When the  plus sign occurs the external potential affects the center mass motion, while in the opposite case the external fields influence on the relative motion. 

We note that the demand given by Eq.(\ref{factorizing}) takes place particularly provided that $F_{\mu \alpha}$ is the tensor of homogeneous electromagnetic fields.

\section{Two  spin-$\frac{1}{2}$ particles in external fields}
\label{sec:3}

We consider   spin-$\frac{1}{2}$ particles with    masses $m_1$ and $m_2$ which interact each other  by means of   vector  and  scalar  fields , which are $A_{int}^\nu (x)$ and $S_{int}(x)$, respectively. These particles also are  in the external vector $A_{ext}^\nu (x)$. Such a two-body system is described by the 16-component spinor which each component  $\varphi_i  (\varepsilon_w,  \bb r)$ is governed by the Klein-Gordon equation (\ref{eq14me-int-3}). Provided that the interaction and external potentials satisfy Eqs.(\ref{restric-1}) (\ref{factorizing}), respectively, these equation is simlified, and  $\varphi_i  (\varepsilon_w, \bb r)$ are governed by Eq.(\ref{eq14me-int-7})

\begin{eqnarray}\label{eq14me-int-7f}
&&\bigg\{(\varepsilon_w + {\cal A}^0 (\bb x))^2+(\bb p +{\cal \bb  A} (\bb x))^2 - (m_w +{\cal S} (\bb x))^2\bigg\}\varphi_i (\varepsilon_w, \bb x)=0. 
\end{eqnarray}

Following the standard way we rewrite this equation in a  form

\begin{eqnarray}\label{eq29}
&&{\hat p}_\nu g^{\nu \mu 
} {\hat p}_\mu  \varphi_i  ( \bb x) =0,
\end{eqnarray}
where we introduce ${\hat p}_\nu =\Big(\sqrt{(\varepsilon_w + {\cal A}^0 (\bb x))^2 -(m_w +{\cal S} (\bb x))^2},-({{ \bb p}}+{\cal \bb  A} (\bb x))\Big)$; and the metric tensor $ g^{\nu \mu 
}$ which non-equal to zeroth diagonal componenets are ${\rm diag} = (1,-1.-1.-1)$.

To obtain the motion equation in the Dirac-like form   we express the metric tensor in terms of the $\gamma$-matrices of  the $(16 \times 16)$ dimension.
The relations (\ref{eq11}) which  introduce new variables  imply going to a new basis in the $\gamma$-matrix space so that the new and old  $\gamma$-matrices are related to each other by means of equations

\begin{eqnarray}\label{eq28}
&&{\hat \Gamma} =\mu_1  \stackrel {(1)} {\gamma}\otimes \stackrel {(2)} {I}+\mu_2  \stackrel {(2)} {\gamma}\otimes \stackrel {(1)} {I},\nonumber \\
&& {\hat{  \gamma}}=  \stackrel {(2)} {{ \gamma}}\otimes \stackrel {(1)} {I}-  \stackrel {(1)} { \gamma}\otimes \stackrel {(2)} {I}
\end{eqnarray}
where the hatted $\hat\gamma$-matrices  form a new basis, the numbers  in brackets which are over a symbol  means   that this matrix acts on the variables of the corresponding particle, $\stackrel {(1,2)} {I}$ is the unit matrix in the space of the first or second  particle, respectively.
These matrices satisfy the following transformation relations (see Appendix B)

\label{eq31}
\begin{eqnarray}\label{eq31}
&&{\hat \Gamma}^0{\hat \Gamma}^0 +{\hat \Gamma}^0{\hat \Gamma}^0= 2 \stackrel {(1+2)} {I};\\
&&{\hat \Gamma}^0 {\hat{  \gamma^a}}+ {\hat{  \gamma^a}}{\hat \Gamma}^0=0, ~~a=1,2,3;\\
&&{\hat{ \gamma^a}}{\hat{ \gamma^b}} +{\hat{ \gamma^b}}{\hat{ \gamma^a}}=-4 \delta^{ab} \stackrel {(1+2)} {I}+\nonumber \\
&&( \sigma_1^a \sigma_2^b+\sigma_2^b \sigma_1^a+ \sigma_2^a \sigma_1^b+\sigma_1^b \sigma_2^a) \stackrel {(1+2)} {I}, 
\end{eqnarray}
where  $\bb s_1$ and $\bb s_2$ denote  particle spins, $\delta^{ab}$ is the Kronecker symbol, $\stackrel {(1+2)} {I}$  is the $16\times 16$ unit matrix. Such transformation relations allow us to write down the metric tensor in terms of  new  $\gamma$- matrices as follows

\label{eq32}
\begin{eqnarray}\label{eq32}
&&\Gamma^\nu \Gamma^\mu +\Gamma^\nu \Gamma^\mu= 2 g^{\nu \mu } +\nonumber \\
&&(\sigma_1^a \sigma_2^b+\sigma_2^b \sigma_1^a+ \sigma_2^a \sigma_1^b+\sigma_1^b \sigma_2^a+4\delta^{ab}) \stackrel {(1+2)/2} {I}, \\
&&\Gamma^\nu = \Big({\hat \Gamma}^0, \frac{{\hat{ \gamma^a}}}{2}\Big)\equiv ({ \Gamma}^0, {{\bb \Gamma}} ) 
\end{eqnarray}

Taking $g^{\nu \mu }$ from Eq.(\ref {eq32}) and substituting it into  Eq.(\ref{eq29}), we obtain the equation

\begin{eqnarray}\label{eq33ab}
&&\Big( \Gamma^0 [(\varepsilon_w + {\cal A}^0 (\bb x))^2 -(m_w +{\cal S} (\bb x))^2]^{1/2}\Big)^2 -  \nonumber \\
&&(\bb \Gamma  {\hat  {\bb p}})^2 \Big)\Psi ( E_w, \bb r) = [\frac{1}{2}(\bb \sigma_1 +\bb \sigma_2)  {\hat {\bb p}} ]^2 \Psi (\varepsilon_w, \bb x) ,
\end{eqnarray}
where $\Psi (\varepsilon_w, \bb r)$ is the 16-component spinor mentioned above in the current section.
 Eq.(\ref{eq33ab})  leads to the Dirac-like form of the motion equations for this spinor

\begin{eqnarray}\label{eq33a}
&& \Gamma_0 [(\varepsilon_w + {\cal A}^0 (\bb x))^2 -(m_w +{\cal S} (\bb x))^2]^{1/2}\Psi (\varepsilon_w, \bb x) =\nonumber \\
&& (\bb \Gamma + \bb S) {\hat {\bb p}}\Psi (\varepsilon_w, \bb x) ,
\end{eqnarray}

\begin{eqnarray}\label{eq33abcd}
&& \Gamma_0 [(\varepsilon_w + {\cal A}^0 (\bb x))^2 -(m_w +{\cal S} (\bb x))^2]^{1/2}\Psi (\varepsilon_w, \bb x) =\nonumber \\
&& (\bb \Gamma - \bb S) {\hat {\bb p}} \Psi (\varepsilon_w, \bb x),
\end{eqnarray}
where $\bb p = -i \nabla$; $ {\hat {\bb p}} ={{ \bb p}}+{\cal \bb  A} (\bb x)$ , $\bb S= (\bb \sigma_1 +\bb \sigma_2)/2$ in Eqs.(\ref{eq33a}), (\ref{eq33abcd}) is assumed to be multiplied by $\stackrel {(1+2)} {I}$. It is very important to mention that  such a structure of  the  $\gamma^\mu$ matrices defined by Eq.(\ref{eq28}) leads to decoupling (see, also Appendix B)  the motion equation (\ref{eq33ab}) in four independent equation as compared with the  previous researches\cite{Sad86-1,Sad86-2,Sad89,Cra84,Cra87,Cra90,Cra01,Cra03,Cra10}.

Assuming the  Eq.(\ref{eq33a}) is the leading,  let us establish the physical sense of  Eqs.(\ref{eq33abcd}). To clarify it we change the numbers of particles in the two-body system. According to Eq. (\ref{eq28}) this  procedure  consists in   changing $\bb p \to -\bb p,~\bb \Gamma \to -\bb \Gamma$. Making  the changes mentioned above in  Eq.(\ref{eq33a}), we obtain the equation 
which coincides with Eq.(\ref{eq33abcd}). Thus,  choosing the leading equation  among  Eqs.(\ref{eq33a}) and Eq.(\ref{eq33abcd}) means fixation of particle  numbering.

Provided that  $\Psi (\varepsilon_w, \bb x)$ is governed by Eq.(\ref{eq33a}),  the direct calculations give  that the Dirac conjugated field  ${\bar  \Psi} (\varepsilon_w, \bb x) = {  \Psi}^\dag  (\varepsilon_w, \bb x) \Gamma^0$ satisfies an equation

\begin{eqnarray}\label{eq33aa}
&&{\bar  \Psi}   (\varepsilon_w, \bb x) \Gamma_0 [(\varepsilon_w + {\cal A}^0 (\bb x))^2 -(m_w +{\cal S} (\bb x))^2]^{1/2} =\nonumber \\
&&-{\bar  \Psi}  (\varepsilon_w, \bb x)(\bb \Gamma + \bb S) {\hat {{\overleftarrow {\bb p}}}}.
\end{eqnarray}

 \subsection{Lagrangian and current of fermions in the center-mass and relative motion variables}

 Eqs.(\ref{eq33a}),  (\ref{eq33aa}) allow us to derive the  Lagrangian governing the dynamics of the two-body system in an external field which is 

\begin{eqnarray}\label{eq33aaa}
&&{\frak L} =\frac{1}{2}  {\bar  \Psi} (\varepsilon_w, \bb x)\Bigg\{ \Big( \Gamma_0 [(\varepsilon_w + {\cal A}^0 (\bb x))^2 -(m_w +{\cal S} (\bb x))^2]^{1/2} -\nonumber \\
&& (\bb \Gamma + \bb S) {\hat {\bb p}} \Big)+ \Big( \Gamma_0 [(\varepsilon_w + {\cal A}^0 (\bb x))^2 -(m_w +{\cal S} (\bb x))^2]^{1/2} +\nonumber \\
&& (\bb \Gamma + \bb S) {\hat {{\overleftarrow {\bb p}}}} \Big)\Bigg\}\Psi (\varepsilon_w, \bb x)
\end{eqnarray}

Such a Lagrangian generates the  current

\begin{eqnarray}\label{eq35a}
&& {\bb J} ( E_w, \bb r) =  {\bar \Psi (\varepsilon_w, \bb x)}(\bb \Gamma +\bb S)     \Psi (\varepsilon_w, \bb x),
\end{eqnarray}
whose  conservation can be expressed in terms of  the continuity equation.

\begin{eqnarray}\label{eq34a}
&&  \nabla~\bb J  (\varepsilon_w, \bb x)= 0.
\end{eqnarray}

\section{A positronium  in a strong magnetic field}
\label{sec:3}

We apply  the obtained result to  a study of observable  physical phenomena,  by  considering a positronium in a strong uniform magnetic field. Such a problem is important   in the astrophysics when  the electromagnetic signals coming from the astrophysical objects such as pulsars or neutron stars\cite{Bro,Bus},  whose energy is about $0.511 MeV$, likely  correspond to the electromagnetic  decay of the positronium  ground state  into two photons.

We study  positronium states in a magnetic field in the Breit frame. Then, $\varepsilon_w= E_w$ in  left-hand side of Eq.(\ref{eq33a}). After that we  take the scalar potential ${\cal S}(\bb r)=0$ therein,  and direct the magnetic field $\bb B$ along the $OZ$ axis.  Further, we go to the second order equitation,  by  acting  on Eq.(\ref{eq33a}) with the left-hand side operator of Eq.(\ref{eq33abcd}). As a result, we get

\begin{eqnarray}\label{eq34}
&&\Big(\triangle +\frac{2(4 \pi \alpha) E_w }{r}+\frac{(4 \pi \alpha)^2 }{r^2}+i\sqrt{\pi \alpha}\nabla (\bb B\times \bb \rho) -\nonumber \\
&&4\pi \alpha  B^2 \rho^2 -m_w^2 +E_w^2\Big)\begin{pmatrix} 1 &0 \\
0 & 1\end{pmatrix}\Psi ( E_w, \bb r)\nonumber \\
&&= - \sqrt{\pi \alpha} 
\begin{pmatrix} \sigma_{z,1} + \sigma_{z,2} &0\\
0  & \sigma_{z,1} + \sigma_{z,2} \end{pmatrix}  \Psi ( E_w, \bb r), 
\end{eqnarray}
where $\alpha = e^2 /4\pi$ is the fine structure constant; $\sigma_{z(1,2)}$ are the Pauli matrices acting on the spin variables of the first (an electron) and the second (a positron) particle. In obtaining the latest equation we have taken

\begin{eqnarray}\label{eq35}
&&A^\mu_{int} = ( A^0_{int}, 0)  = (-\frac{4 \pi \alpha}{r},0); ~{\cal A}_{ext} = \frac{1}{2}(\bb B\times \bb r).
\end{eqnarray}

Eq.(\ref{eq34}) contains two diagonal matrices that allows us to go to two-component functions $ \Psi ( E_w, \bb r)$

\begin{eqnarray}\label{eq34a}
&&\Big(\triangle +\frac{2(4 \pi \alpha) E_w }{r}+\frac{(4 \pi \alpha)^2 }{r^2}+i\sqrt{\pi \alpha}\nabla (\bb B\times \bb r) -\nonumber \\
&&4\pi \alpha  B^2 \rho^2 -m_w^2 +E_w^2\Big)\Psi ( E_w, \bb r)\nonumber \\
&&= - \sqrt{\pi \alpha}( \sigma_{z,1} + \sigma_{z,2} ) \Psi ( E_w, \bb r).
\end{eqnarray}

We study the low energy  states ($S=S_z=0$) of  a positronium, which are differed by the spin number $S$,  in a strong magnetic field $\bb B$ when the magnetic length $ a = (2\sqrt{\pi \alpha} B)^{-1/2}$ is  much smaller   than the positronium Bohr radius $a_B = 1 /2\pi \alpha m_e $

\begin{eqnarray}\label{str-mf}
&& ~~~~~~~~~~~~~~~~~~~~~~~a\ll a_B, 
\end{eqnarray}
 where   $m_e$ is the electron mass. 

However,  the ground state ($S=S_z=0$) turn out to be mixed\cite{LL4} with the triplet spin state $(S=1, S_z=0)$ due to the right-hand side terms in Eq.(\ref{eq34a}). Therefore, we look for solution of this equation as follows

\begin{eqnarray}\label{eq36d}
&& \Psi ( E_w, \bb r) = \frac{\psi_{m=0,n=0} (\bb \rho)}{\sqrt 2} \sum\limits_{S=0}^1\Bigg( \begin{pmatrix} 1 \\
0  \end{pmatrix}_1   \begin{pmatrix} 0 \\
1  \end{pmatrix}_2 +(-1)^{S+1} \begin{pmatrix} 0 \\
1 \end{pmatrix}_1   \begin{pmatrix} 1 \\
0  \end{pmatrix}_2
\Bigg) \psi_{S} (z),
\end{eqnarray}
where the indexes mean the number of a particle; $\psi_{m=0,n=0} (\bb \rho)$ is the wave function governing  the transverse motion of a particle in the ground state in a magnetic field\cite{LL3}, which is equal to

\begin{eqnarray}\label{eq36}
&&\psi_{m=0,n=0} (\bb \rho) =\frac{1}{a\sqrt{2 \pi}} \exp{\left(-\frac{\rho^2}{4 a^2}\right)} .
\end{eqnarray}
Here  the cylindrical coordinates $\bb r = \bb \rho +\bb e_z z$ are introduced, $m$  is the projection of an angular momentum along the $\bb B$ direction,  $n$ is a radial quantum number.  
Substituting $\Psi ( E_w, \bb r)$ given by  Eq.(\ref{eq36d}) into Eq.(\ref{eq34a}), we derive

\begin{eqnarray}\label{eq36ad}
&&\Big(\frac{d^2}{d\xi^2} +  2^{1/2}\sqrt{\pi}(4 \pi \alpha)E_w a~ e^{\xi^2/2} erfc (|\xi| /\sqrt 2) +\nonumber \\
&&\frac{ (4 \pi \alpha)^2 e^{\xi^2/2}}{2} E_1 (\xi^2/2) -\varepsilon^2_S \Big)\psi_S (\xi)= - 2\sqrt{\pi \alpha}\psi_{1-S} (\xi), \nonumber \\
\end{eqnarray}
where the introduced   $erfc (x)$ is the complementary error function, $E_1(x)$ is the integral exponent\cite{Gra}. We have also denoted in Eq.(\ref{eq36ad})

\begin{eqnarray}\label{eq39}
&&z /a \equiv \xi,~~  ( m_w^2 -E_w^2)a^2+1+(-1)^{S+1} \equiv \varepsilon_S^2.
\end{eqnarray}

Since  the magnetic field is strong  the particle\footnote{The compound particle - a positronium - is meant here.}  is weakly bound with respect to  a motion along the $OZ$ axis as compared with the transverse motion. Besides that, the term

\begin{eqnarray}\label{eq36aad}
&&-U(\xi) \equiv \Bigg(  2^{1/2}\sqrt{\pi}(4 \pi \alpha) E_w a e^{\xi^2/2} erfc (|\xi| /\sqrt 2) +\nonumber \\
&&\frac{ (4 \pi \alpha)^2 e^{\xi^2/2}}{2} E_1 (\xi^2/2)\Bigg) 
\end{eqnarray}
  is weakly singular since $U(\xi) \propto -1/\sqrt \xi$ at small $\xi \to 0$. Therefore, we change the exact potential $U(\xi)$ by its approximate form
 
 \begin{eqnarray}\label{eq36aad}
&&U(\xi)\simeq \delta (\xi)  \int\limits_{-\infty}^{+\infty}\psi^\ast_S (\xi) U(\xi)\psi_S (\xi) d \xi , 
\end{eqnarray}
 where $\psi_S (\xi)$ satisfies an equation

\begin{eqnarray}\label{eq36aa}
&&\Big(\frac{d^2}{d\xi^2} - \delta (\xi)(  \int\limits_{-\infty}^{+\infty}\psi^\ast_S (\xi') U(\xi')\psi_S (\xi')d\xi')-\varepsilon^2_S\Big)\psi_S (\xi) =\nonumber \\
&& - 2\sqrt{\pi \alpha}\psi_{1-S} (\xi), \nonumber \\
\end{eqnarray}
We substitute $\psi_S (\xi)$,  given by the formula 

\begin{eqnarray}\label{eq40}
\psi_S (\xi) =\sqrt{ k_S} \exp ( - k_S |\xi|),
\end{eqnarray}
into Eq.(\ref{eq36aa}),  where $k_S=\sqrt \varepsilon_S$ is a positive constant. As a result,  we derive  the  equation to obtain $k_S$

\begin{eqnarray}\label{eq41}
&&k_S  =  {(4 \pi \alpha)^2}\int\limits_0^\infty  d \xi \exp{({\xi^2/2}-2k_S \xi)} E_1 (\xi^2/2) + \nonumber \\
&&2^{3/2}\sqrt{\pi}(4 \pi \alpha) E_w a \int\limits_0^\infty  d \xi \exp{({\xi^2/2}-2k_S \xi)}erfc (\xi /\sqrt 2)\nonumber \\
\end{eqnarray}
The validity limit of the derived equation is $k_S\ll 1$.  At such $k_S$  the second integral is approximately  $\pi^{3/2}/\sqrt 2$ \cite{Gra}, whereas the first integral can  be  approximately taken  to be equal to  $\ln (a_B/a)/\sqrt \pi$\cite{LL3} after  the dimensional  regularization.  

As a result, we find

\begin{eqnarray}\label{eq42}
&& k_S  =  2^{3/2}(4 \pi \alpha) E_w a \ln (a_B/a)+(4 \pi \alpha)^2\frac{\pi^{3/2}}{\sqrt 2}.
\end{eqnarray}

In the non-relativistic  case $m_e a \gg\alpha$  we obtain  $k_S  =  2^{3/2}(4 \pi \alpha/3) m_e a\ln (a_B/a)$.  Taking into account  of  Eq.(\ref{eq42}), we derive
\begin{eqnarray}\label{eq43}
&& E_w =\frac{ m_e}{2} \Big( 1 + \frac{2 (1 +(-1)^{S+1})}{ (m_e a)^2 } - \nonumber \\
&&2 (4 \pi \alpha/3)^2 \ln^2 (a_B/a)\Big),
\end{eqnarray}
that relates to the results obtained earlier\cite{Ell}.

In the opposite limiting situation $m_e a \ll \alpha$, but provided that  the vacuum polarization effects  in a strong magnetic field are absent\cite{LL4}, we get  $k_S =9(2 \pi)^{7/2} \alpha^2$, that gives 
\begin{eqnarray}\label{eq44}
&& E_w \simeq \frac{ 1}{a^2} ( (1 +(-1)^{S+1}) -  (2 \pi)^7  \alpha^4/162).
\end{eqnarray}
The obtained $E_w$ sufficiently differs from the  results has been derived above in the non-relativistic case. Eqs.(\ref{eq43}), (\ref{eq44}) show the stronger a magnetic field , the less   the positronium level  depth  under the bottom of the main Landau zone.

\subsection{Decay of the positronium ground state in a strong magnetic field}
\label{sec:3}

The decay width $\Gamma$ of a particle is proportional  to\cite{Green}
\begin{eqnarray}\label{eq45}
&& \Gamma \simeq \frac{|\psi (0)|^2}{M^2}, 
\end{eqnarray}
where $\psi (0)$ and $M$ are the wave function and mass of  the decaying particle, respectively.

According to Eqs.(\ref{eq36}), (\ref{eq40})  the squared wave function module $|\psi (0)|^2$, depending on the  absence or presence  of a magnetic field, is of the order of 
\begin{eqnarray}\label{eq46a}
 &&|\psi_{B=0} (0)|^2 \sim a_B^{-3}~~~~~~~~~~|\psi_{B\ne0} (0)|^2 \sim\frac{ 1}{a_B a^2}. 
\end{eqnarray}
 
The estimations given by Eq.(\ref{eq46a}) show that a strong magnetic field leads to sufficient increasing, in $(a_B/a)^2  \gg 1$ times,  the decay width  as  compared with the case  $B=0$. Such a behavior of the decay width in a magnetic field  is a  result of the additional compressing  the electron-positron bound  state by a magnetic field in the plane which is perpendicular to the ${\bf B}$-direction.

\section{ Conclusion}
\label{sec:5}
In terms of the principle of   least action we study the open relativistic two-body  problem, when two  interacting particles also undergo  effects of  external potentials. Starting from  the single particle action integral  for a  spinless particle we derive the Lagrangian governing the dynamics of both the two-body  system  of scalar particles  and each of the   16 components of a spinor determoning  the  two-fermion system in external fields, beyond the consideration in the Breit frame. The motion equations,  generated such a Lagrangian,   take into account  the modification of  Newton's third law due to  in external fields, and have, in the general case, the  operator structure in the variables of the relative momentum and the momentum of the center mass of the  two-body system.     Provided that Newton's third law takes place  the derived equations are simplified, and turn out in  a good relation to the equations have been obtained earlier in the constraint dynamic approach \cite{Cra84,Cra87,Cra90,Cra01,Cra03,Cra10} in the absence of external fields.
 Following the standard way, based on  using the $\gamma$-matrix basis,  the Klein-Gordon equation governing the evolution of  each of the 16 components of the relativistic spinor is reduced to the Dirac-like equations. The  new   $\gamma$-matrix basis,  when the derived Dirac-like equations turn out  to be decoupled  in four independent coupled  equations which determine the dynamics of four-component spinors,  is found . Based on  the derived Dirac-like equations  we obtain the  Lagrangian  which  generates the conserving fermion current.  In the case of  the pure   QED the developed approach has been  applied to  studying a  positronium  in a strong magnetic field. The positronium low energy  states are studied, taking into account spin-mixing effect\cite{LL4}.  It is shown that strong magnetic fields result in essentially reconstructing the spectrum of the positronium states and decreasing the life-time of the positronium ground state.

\appendix{}

\section{The transformation relation for $\Gamma$- matrices}

According to  Eqs.(\ref{eq28}) we have

\begin{eqnarray}
&&{\hat \Gamma} =\mu_1  \stackrel {(1)} {\gamma}\otimes \stackrel {(2)} {I}+\mu_2  \stackrel {(2)} {\gamma}\otimes \stackrel {(1)} {I},\nonumber \\
&& {\hat{ \bb \gamma}}=  \stackrel {(2)} {{\bb \gamma}}\otimes \stackrel {(1)} {I}-  \stackrel {(1)} {\bb \gamma}\otimes \stackrel {(2)} {I}.
\end{eqnarray}

The matrices ${\hat \Gamma}$  and  $\hat{ \bb \gamma}$  are explicitly given by formulas

\begin{eqnarray}
&&{\hat \Gamma}^0= (\mu_1  \stackrel {(1)} {\gamma^0}\otimes \stackrel {(2)} {I}+\mu_2  \stackrel {(2)} {\gamma^0}\otimes \stackrel {(1)} {I})= \mu_1 \begin{pmatrix}
\stackrel {(1)} {\gamma^0} & 0 & 0&0\\
0&  \stackrel {(1)} {\gamma^0} & 0 &0\\
0&  0 & \stackrel {(1)} {\gamma^0} &0\\
0&  0 & 0 &\stackrel {(1)} {\gamma^0}\\
\end{pmatrix}+\mu_2  \begin{pmatrix}
\stackrel {(2)} {\gamma^0} & 0 & 0&0\\
0&  \stackrel {(2)} {\gamma^0} & 0 &0\\
0&  0 & \stackrel {(2)} {\gamma^0} &0\\
0&  0 & 0 &\stackrel {(2)} {\gamma^0}\\
\end{pmatrix} 
\end{eqnarray}

\begin{eqnarray}
&&{\hat{\bb  \gamma}}= \begin{pmatrix}
\stackrel {(2)} {\bb \gamma} & 0 & 0&0\\
0&  \stackrel {(2)} {\bb \gamma} & 0 &0\\
0&  0 & \stackrel {(2)} {\bb \gamma} &0\\
0&  0 & 0 &\stackrel {(2)} {\bb \gamma}\\
\end{pmatrix} -\begin{pmatrix}
\stackrel {(1)} {\bb \gamma} & 0 & 0&0\\
0&  \stackrel {(1)} {\bb \gamma} & 0 &0\\
0&  0 & \stackrel {(1)} {\bb \gamma} &0\\
0&  0 & 0 &\stackrel {(1)} {\bb \gamma}\\
\end{pmatrix},
\end{eqnarray}
where a number  in the  brackets which is  over $\gamma$ means that this matrix acts on the variable of the corresponding particle, $\stackrel {(a)} {I}$ is the unit matrix in the space of the $a$ particle. Then, we obtain

\begin{eqnarray}
&&({\hat \Gamma}^0{\hat \Gamma}_0 +{\hat \Gamma}^0{\hat \Gamma}_0) =2 {\hat \Gamma}^0{\hat \Gamma}_0= 2(\mu_1  \stackrel {(1)} {\gamma^0}\otimes \stackrel {(2)} {I}+\mu_2  \stackrel {(2)} {\gamma^0}\otimes \stackrel {(1)} {I})\times \nonumber \\ &&(\mu_1  \stackrel {(1)} {\gamma^0}\otimes \stackrel {(2)} {I}+\mu_2  \stackrel {(2)} {\gamma^0}\otimes \stackrel {(1)} {I})=2(\mu_1^2  \stackrel {(1+2)} {I}+\mu_2^2  \stackrel {(1+2)} {I}+\nonumber \\
 &&\mu_1 \mu_2 ( \stackrel {(2)} {\gamma^0}\otimes \stackrel {(1)} {I} \cdot \stackrel {(1)} {\gamma^0}\otimes \stackrel {(2)} {I}+\stackrel {(1)} {\gamma^0}\otimes \stackrel {(2)} {I} \cdot \stackrel {(2)} {\gamma^0}\otimes \stackrel {(1)} {I}))\nonumber \\
&&=2(\mu_1^2  \stackrel {(1+2)} {I}+\mu_2^2  \stackrel {(1+2)} {I}+2\mu_1 \mu_2 ( \stackrel {(1+2)} {\gamma^0}\cdot \stackrel {(1+2)} {\gamma^0}))= 2\stackrel {(1+2)} {I}, \nonumber \\ 
\end{eqnarray}
where $\stackrel {(1+2)} {I}$ is the $(16\times 16)$ unit matrix.

Calculation of the anti-commutator $[{\hat \Gamma}^0, {\hat{\bb  \gamma}}]_+$ leads to the following results

\begin{eqnarray}
&&{\hat \Gamma}^0 {\hat{\bb  \gamma}}+ {\hat{\bb  \gamma}}{\hat \Gamma}^0=(\mu_1 \begin{pmatrix}
\stackrel {(1)} {\gamma^0} & 0 & 0&0\\
0&  \stackrel {(1)} {\gamma^0} & 0 &0\\
0&  0 & \stackrel {(1)} {\gamma^0} &0\\
0&  0 & 0 &\stackrel {(1)} {\gamma^0}\\
\end{pmatrix}+\mu_2  \begin{pmatrix}
\stackrel {(2)} {\gamma^0} & 0 & 0&0\\
0&  \stackrel {(2)} {\gamma^0} & 0 &0\\
0&  0 & \stackrel {(2)} {\gamma^0} &0\\
0&  0 & 0 &\stackrel {(2)} {\gamma^0}\\
\end{pmatrix}  )\times \nonumber \\ 
&&(\begin{pmatrix}
\stackrel {(2)} {\bb \gamma} & 0 & 0&0\\
0&  \stackrel {(2)} {\bb \gamma} & 0 &0\\
0&  0 & \stackrel {(2)} {\bb \gamma} &0\\
0&  0 & 0 &\stackrel {(2)} {\bb \gamma}\\
\end{pmatrix} -\begin{pmatrix}
\stackrel {(1)} {\bb \gamma} & 0 & 0&0\\
0&  \stackrel {(1)} {\bb \gamma} & 0 &0\\
0&  0 & \stackrel {(1)} {\bb \gamma} &0\\
0&  0 & 0 &\stackrel {(1)} {\bb \gamma}\\
\end{pmatrix} )+(\begin{pmatrix}
\stackrel {(2)} {\bb \gamma} & 0 & 0&0\\
0&  \stackrel {(2)} {\bb \gamma} & 0 &0\\
0&  0 & \stackrel {(2)} {\bb \gamma} &0\\
0&  0 & 0 &\stackrel {(2)} {\bb \gamma}\\
\end{pmatrix} -\begin{pmatrix}
\stackrel {(1)} {\bb \gamma} & 0 & 0&0\\
0&  \stackrel {(1)} {\bb \gamma} & 0 &0\\
0&  0 & \stackrel {(1)} {\bb \gamma} &0\\
0&  0 & 0 &\stackrel {(1)} {\bb \gamma}\\
\end{pmatrix} )\times \nonumber \\ 
&&(\mu_1 \begin{pmatrix}
\stackrel {(1)} {\gamma^0} & 0 & 0&0\\
0&  \stackrel {(1)} {\gamma^0} & 0 &0\\
0&  0 & \stackrel {(1)} {\gamma^0} &0\\
0&  0 & 0 &\stackrel {(1)} {\gamma^0}\\
\end{pmatrix}+\mu_2  \begin{pmatrix}
\stackrel {(2)} {\gamma^0} & 0 & 0&0\\
0&  \stackrel {(2)} {\gamma^0} & 0 &0\\
0&  0 & \stackrel {(2)} {\gamma^0} &0\\
0&  0 & 0 &\stackrel {(2)} {\gamma^0}\\
\end{pmatrix}  )
\end{eqnarray}

\begin{eqnarray}
&&{\hat \Gamma}^0 {\hat{\bb  \gamma}}+ {\hat{\bb  \gamma}}{\hat \Gamma}^0=\nonumber \\
&& \mu_1 (\begin{pmatrix}
\stackrel {(1)} {\gamma^0} \stackrel {(2)} {\bb \gamma} & 0 & 0&0\\
0&  \stackrel {(1)} {\gamma^0} \stackrel {(2)} {\bb \gamma} & 0 &0\\
0&  0 & \stackrel {(1)} {\gamma^0}\stackrel {(2)} {\bb \gamma}  &0\\
0&  0 & 0 &\stackrel {(1)} {\gamma^0}\stackrel {(2)} {\bb \gamma} \\
\end{pmatrix}- \begin{pmatrix}
\stackrel {(1)} {\gamma^0} \stackrel {(1)} {\bb \gamma} & 0 & 0&0\\
0&  \stackrel {(1)} {\gamma^0} \stackrel {(1)} {\bb \gamma} & 0 &0\\
0&  0 & \stackrel {(1)} {\gamma^0}\stackrel {(1)} {\bb \gamma}  &0\\
0&  0 & 0 &\stackrel {(1)} {\gamma^0}\stackrel {(1} {\bb \gamma} \\
\end{pmatrix}+\begin{pmatrix}
\stackrel {(2)} {\bb \gamma} \stackrel {(1)} {\gamma^0}& 0 & 0&0\\
0&  \stackrel {(2)} {\bb \gamma} \stackrel {(1)} {\gamma^0}& 0 &0\\
0&  0 & \stackrel {(2)} {\bb \gamma} \stackrel {(1)} {\gamma^0}&0\\
0&  0 & 0 &\stackrel {(2)} {\bb \gamma}\stackrel {(1)} {\gamma^0}\\
\end{pmatrix})\nonumber \\ 
&&-\begin{pmatrix}
\stackrel {(1)} {\bb \gamma} \stackrel {(1)} {\gamma^0}& 0 & 0&0\\
0&  \stackrel {(1)} {\bb \gamma} \stackrel {(1)} {\gamma^0}& 0 &0\\
0&  0 & \stackrel {(1)} {\bb \gamma} \stackrel {(1)} {\gamma^0}&0\\
0&  0 & 0 &\stackrel {(1)} {\bb \gamma}\stackrel {(1)} {\gamma^0}\\
\end{pmatrix})
+ \mu_2 (\begin{pmatrix}
\stackrel {(2)} {\gamma^0} \stackrel {(2)} {\bb \gamma} & 0 & 0&0\\
0&  \stackrel {(2)} {\gamma^0} \stackrel {(2)} {\bb \gamma} & 0 &0\\
0&  0 & \stackrel {(2)} {\gamma^0}\stackrel {(2)} {\bb \gamma}  &0\\
0&  0 & 0 &\stackrel {(2)} {\gamma^0}\stackrel {(2)} {\bb \gamma} \\
\end{pmatrix}-\begin{pmatrix}
\stackrel {(2)} {\gamma^0} \stackrel {(1)} {\bb \gamma} & 0 & 0&0\\
0&  \stackrel {(2)} {\gamma^0} \stackrel {(1)} {\bb \gamma} & 0 &0\\
0&  0 & \stackrel {(2)} {\gamma^0}\stackrel {(1)} {\bb \gamma}  &0\\
0&  0 & 0 &\stackrel {(2)} {\gamma^0}\stackrel {(1} {\bb \gamma} \\
\end{pmatrix}
\nonumber \\&&+\begin{pmatrix}
\stackrel {(2)} {\bb \gamma} \stackrel {(2)} {\gamma^0}& 0 & 0&0\\
0&  \stackrel {(2)} {\bb \gamma} \stackrel {(2)} {\gamma^0}& 0 &0\\
0&  0 & \stackrel {(2)} {\bb \gamma} \stackrel {(2)} {\gamma^0}&0\\
0&  0 & 0 &\stackrel {(2)} {\bb \gamma}\stackrel {(2)} {\gamma^0}\\
\end{pmatrix})-\begin{pmatrix}
\stackrel {(1)} {\bb \gamma} \stackrel {(2)} {\gamma^0}& 0 & 0&0\\
0&  \stackrel {(1)} {\bb \gamma} \stackrel {(2)} {\gamma^0}& 0 &0\\
0&  0 & \stackrel {(1)} {\bb \gamma} \stackrel {2)} {\gamma^0}&0\\
0&  0 & 0 &\stackrel {(1)} {\bb \gamma}\stackrel {(2)} {\gamma^0}\\
\end{pmatrix}).\nonumber \\ \nonumber \\
\end{eqnarray}

Then, we  finally  derive

\begin{eqnarray}
&&{\hat \Gamma}^0 {\hat{\bb  \gamma}}+ {\hat{\bb  \gamma}}{\hat \Gamma}^0=0.
\end{eqnarray}

Let us go to the anti-commutator $ [{\hat{\bb \gamma^a},}{\hat{\bb \gamma^b}}]_+$

\begin{eqnarray}
&&({\hat{\bb \gamma^a}}{\hat{\bb \gamma^b}} +{\hat{\bb \gamma^b}}{\hat{\bb \gamma^a}}) =
 (\stackrel {(2)} {{\bb \gamma^a}}\otimes \stackrel {(1)} {I}-  \stackrel {(1)} {\bb \gamma^a}\otimes \stackrel {(2)} {I})(\stackrel {(2)} {{\bb \gamma^b}}\otimes \stackrel {(1)} {I}-  \stackrel {(1)} {\bb \gamma^b}\otimes \stackrel {(2)} {I})\nonumber \\
 +&&(\stackrel {(2)} {{\bb \gamma^b}}\otimes \stackrel {(1)} {I}-  \stackrel {(1)} {\bb \gamma^b}\otimes \stackrel {(2)} {I})(\stackrel {(2)} {{\bb \gamma^a}}\otimes \stackrel {(1)} {I}-  \stackrel {(1)} {\bb \gamma^a}\otimes \stackrel {(2)} {I}).
\end{eqnarray}

Substituting the explicit form of ${\hat{\bb \gamma^a}}$ we get
\begin{eqnarray}
&&({\hat{\bb \gamma^a}}{\hat{\bb \gamma^b}} +{\hat{\bb \gamma^b}}{\hat{\bb \gamma^a}}) =(\begin{pmatrix}
\stackrel {(2)} {\bb \gamma^a} & 0 & 0&0\\
0&  \stackrel {(2)} {\bb \gamma^a} & 0 &0\\
0&  0 & \stackrel {(2)} {\bb \gamma^a} &0\\
0&  0 & 0 &\stackrel {(2)} {\bb \gamma^a}\\
\end{pmatrix} -\begin{pmatrix}
\stackrel {(1)} {\bb \gamma^a} & 0 & 0&0\\
0&  \stackrel {(1)} {\bb \gamma^a} & 0 &0\\
0&  0 & \stackrel {(1)} {\bb \gamma^a} &0\\
0&  0 & 0 &\stackrel {(1)} {\bb \gamma^a}\\
\end{pmatrix} )\times (\begin{pmatrix}
\stackrel {(2)} {\bb \gamma^b} & 0 & 0&0\\
0&  \stackrel {(2)} {\bb \gamma^b} & 0 &0\\
0&  0 & \stackrel {(2)} {\bb \gamma^b} &0\\
0&  0 & 0 &\stackrel {(2)} {\bb \gamma^b}\\
\end{pmatrix} -\begin{pmatrix}
\stackrel {(1)} {\bb \gamma^b} & 0 & 0&0\\
0&  \stackrel {(1)} {\bb \gamma^b} & 0 &0\\
0&  0 & \stackrel {(1)} {\bb \gamma^b} &0\\
0&  0 & 0 &\stackrel {(1)} {\bb \gamma^b}\\
\end{pmatrix} )\nonumber \\
&&+(\begin{pmatrix}
\stackrel {(2)} {\bb \gamma^b} & 0 & 0&0\\
0&  \stackrel {(2)} {\bb \gamma^b} & 0 &0\\
0&  0 & \stackrel {(2)} {\bb \gamma^b} &0\\
0&  0 & 0 &\stackrel {(2)} {\bb \gamma^b}\\
\end{pmatrix} -\begin{pmatrix}
\stackrel {(1)} {\bb \gamma^b} & 0 & 0&0\\
0&  \stackrel {(1)} {\bb \gamma^b} & 0 &0\\
0&  0 & \stackrel {(1)} {\bb \gamma^b} &0\\
0&  0 & 0 &\stackrel {(1)} {\bb \gamma^b}\\
\end{pmatrix} )\times (\begin{pmatrix}
\stackrel {(2)} {\bb \gamma^a} & 0 & 0&0\\
0&  \stackrel {(2)} {\bb \gamma^a} & 0 &0\\
0&  0 & \stackrel {(2)} {\bb \gamma^a} &0\\
0&  0 & 0 &\stackrel {(2)} {\bb \gamma^a}\\
\end{pmatrix} -\begin{pmatrix}
\stackrel {(1)} {\bb \gamma^a} & 0 & 0&0\\
0&  \stackrel {(1)} {\bb \gamma^a} & 0 &0\\
0&  0 & \stackrel {(1)} {\bb \gamma^a} &0\\
0&  0 & 0 &\stackrel {(1)} {\bb \gamma^a}\\
\end{pmatrix} ).\nonumber \\
\end{eqnarray}

Making obvious transformation we derive

\begin{eqnarray}
&&({\hat{\bb \gamma^a}}{\hat{\bb \gamma^b}} +{\hat{\bb \gamma^b}}{\hat{\bb \gamma^a}}) =4 g^{ab} \stackrel {(1+2)/2} {I}+ (\sigma_1^a \sigma_2^b+\sigma_2^b \sigma_1^a+ \sigma_2^a \sigma_1^b+\sigma_1^b \sigma_2^a) \stackrel {(1+2)/2} {I}\nonumber \\
&&=-4 \delta^{ab} \stackrel {(1+2)} {I}+(\sigma_1^a \sigma_2^b+\sigma_2^b \sigma_1^a+ \sigma_2^a \sigma_1^b+\sigma_1^b \sigma_2^a) \stackrel {(1+2)/2} {I} .\nonumber \\ 
\end{eqnarray}

\end{document}